\documentclass[onecolumn]{aastex63}

\usepackage{natbib} 

\usepackage{graphicx} % Required for inserting images
\usepackage{amssymb}	% Extra maths symbols \usepackage{amsmath}
\usepackage{xcolor} 
\newcommand{\Msol}{\,\mathrm{M}_\odot}

\begin{document}

\title{\sc {  Delayed emission from luminous  blue optical transients in
black-hole  binary systems} } \author[0000-0002-9190-662X]{Davide
Lazzati} \affiliation{Department of Physics, Oregon State University,
301 Weniger Hall, Corvallis, OR 97331, USA}
\author[0000-0002-3635-5677]{Rosalba Perna} \affiliation{Department of
Physics and Astronomy, Stony Brook University, Stony Brook, NY
11794-3800, USA} \author[0000-0002-0786-7307]{Taeho Ryu} \affil{The Max
Planck Institute for Astrophysics, Karl-Schwarzschild-Str. 1, Garching,
85748, Germany} \author[0000-0001-5228-6598]{Katelyn Breivik}
\affil{McWilliams Center for Cosmology and Astrophysics, Department of
Physics, Carnegie Mellon University, Pittsburgh, PA 15213, USA}

\begin{abstract} At least three members of the recently identified class
of fast luminous blue optical transient show evidence of late-time
electromagnetic activity in great excess of what predicted by an
extrapolation of the early time emission. In particular, AT2022tsd
displays fast, bright optical fluctuations approximately a month after
the initial detection. Here, we propose that these transients are
produced by exploding stars in black hole binary systems, and that the
late-time activity is due to the accretion of clumpy  ejecta onto the
companion black hole. We derive the energetics and timescales involved,
compute the emission spectrum, and discuss whether the ensuing emission
is diffused or not in the remnant. We find that this model can explain
the observed range of behaviors for reasonable ranges of the orbital
separation and the ejecta velocity and clumpiness. Close separation and
clumpy, high velocity ejecta result in bright variable emission, as seen
in AT2022tsd. A wider separation and smaller ejecta velocity,
conversely, give rise to fairly constant emission at a lower luminosity.
We suggest that high-cadence, simultaneous, panchromatic monitoring of
future transients should be carried out to better understand the origin
of the late emission and the role of binarity in the diversity of
explosive stellar transients.

\end{abstract}


\section{Introduction} \label{sec:intro}

The last decades have seen the discovery of a variety of new transients,
some of which do not immediately fit already well known phenomena. An
especially interesting new class is that of the  Fast Blue Optical
Transients (FBOTs, e.g. \citealt{Drout2014}), characterized by peak
luminosities $\sim 10^{41}-10^{44}$~erg~s$^{-1}$, blue colors, and rapid
rises, generally $< 10$~days. Among these, especially interesting is the
sub-class of rarer but more extremely luminous events with peak
luminosities at the upper range of $\sim 10^{44}$~erg~s$^{-1}$, which is
often called Luminous Fast Blue Optical Transients (LFBOTs).

Unlike for ordinary supernovae (SNe), the characteristics of the LFBOTs
light curves cannot be readily explained via radioactive decay of
$^{56}$Ni, hence requiring alternative mechanisms for powering their
energetic outputs. Two broad classes of models have been proposed: those
based on the interaction of an outflow with a dense circumstellar
medium, which is shocked and helps convert the flow's kinetic energy
into radiation \citep{Leung2021,Gottlieb2022,Pellegrino2022}, and those
based on a central engine which continues to be active for an extended
period of time. Persistent sources of energy can be of different nature,
from black hole (BH) accretion following the collapse of a supergiant
star \citep{Margutti2019,Perley2019,Quataert2019}, to magnetars
\citep{Prentice2018,Mohan2020,Pasham2022}, to BHs accreting the tidal
disrupted debris of a star \citep{Perley2019,Kuin2019}. Other proposed
models include the merger of a Wolf-Rayet Star with a compact object
companion \citep{Metzger2022}.

The presence of a long-lived source, while hinted in some of the FBOTS,
has however been particularly evident in some of the most extreme
LFBOTs, such as AT2018cow \citep{Perley2019} and AT2022tsd
\citep{Ho2023}, as well as AT2020mrf \citep{Yao2022}. In the case of
At2018cow, following a steep decay of about 200~days, a low-level X-ray
flux with luminosity $\sim 10^{39}$~erg~s$^{-1}$ was observed
\citep{Migliori2024} after more than 1000~days. Spectral analysis
revealed the late emission to be of different origin than the primary
one from the LFBOT. Observations with the {\em Hubble Space Telescope}
in the UV further revealed a source of $\sim 10^{39}$~erg~s$^{-1}$ at
703~days and at 1453~days \citep{Chen2023}, with a mild decay between
the two epochs, which argued against being a stable background, but
rather a late-time association to AT2018cow.

The transient AT2020mrf \citep{Yao2022} had an optical spectrum
displaying a strong resemblance to the one of AT2018cow. Likewise this
transient, it was observed to exhibit late-time X-ray emission, albeit
more than 200 times more intense than for the AT2018cow. In particular,
at 328~days after its start, the {\em Chandra } telescope detected the
source at a level of $L_X\sim 10^{42}$~erg~s$^{-1}$. The emission was
found to be variable on a timescale of $\sim 1$~day.

The AT2022tsd transient has similar evidence for the presence of a
long-lasting, possibly distinct secondary emission component. Starting
at about 26 days after the initial discovery, optical photometry
revealed flaring activity, persisting for about 100 days and with
intensity nearly comparable to that of the original transient, $\sim
10^{44}$~erg~s$^{-1}$ \citep{Ho2023}.

The presence of a secondary, persistent energy source places further
constraints on models for LFBOTs. Here we argue that, in LFBOT models in
which the primary emission is produced as a result of a SN explosion,
the extended emission can be reproduced if the newly formed BH (or
magnetar) has a BH companion formed from a former SN explosion which did
not unbind the binary.

This fraction of survived binaries has been studied both relying on
observations \citep[e.g.,][]{Kochanek+2021,
Neustadt+2021,ByrneFraser2022}, as well as by adopting population
synthesis models (e.g. \citealt{Kochanek+2019}). It was found to range
between a few percent and several tens of percent.

Our model relies on a naturally occurring phenomenon connected to the
second SN explosion: the interaction of the SN ejecta with the remnant
BH of the primary star. The possibility of electromagnetic emission from
the birth of binary BHs was previously discussed by \citet{Kimura2017a}
and \citet{Kimura2017b}. They studied outflow-driven transients in
tidally-locked SNe where the progenitor binary is made up of a
Wolf-Rayet star and a BH. When the second star goes off as a SN, some of
the material is gravitationally captured by the primary BH, forming an
accretion disk and driving a powerful wind. The energy of the wind, in
turn, is deposited in the rest of the ejecta, and radiation diffusively
emerges. Another binary-driven emission scenario was proposed by
\citet{Fryer2014} as a gamma-ray burst (GRB) model: ejecta from the
explosion of a SN rapidly accrete onto a neutron star companion, causing
it to collapse onto a BH. The hyperaccreting conditions of the induced
gravitational collapse would lead to a GRB-like transient.

Here we specifically focus on the LFBOT late-time emission, which in at
least some cases follows after a period of quietness. We suggest that it
is the result of accretion of clumpy ejecta from the SN of the secondary
star onto the primary BH. Unlike traditional models in which the late
emission is produced by the same primary engine, a quiescent phase is
naturally predicted in the case of an accreting companion, correlating
with the time that the SN ejecta take to travel from the exploding star
to the BH companion. Additionally, the optical depth of the ejecta
surrounding the companion is expected to be smaller than if the central
engine were located at the center of the SN.

We specifically focus on a scenario in which the SN ejecta are
inhomogeneous, as found both from SN observations
\citep{Fransson1989,Jerkstrand2011,Abellan2017}, as well as theoretical
models \citep{Wang-Chev2002,Wang2001SNclump,Dessart2018}. In this case,
the resulting accretion is expected to be episodic, possibly producing
flare-like emission. As quantified in this work, the observable
properties of these  transients will mainly depend on the ejecta
properties, such as it mass, velocity, clumping fraction, as well as the
orbital separation and mass of the BH companion.

This paper is organized as follows: first we summarize the observations,
with special emphasis on the late emission of the three sources we study
(Sec.~\ref{sec:at2022tsd}); we then present our analytic model for late
emission due to accretion of ejecta clumps onto the black hole companion
(Sec.~\ref{sec:model}), discussing in detail the expected accretion
rates (Sec.~3.1), the emission mechanisms (Sec.~3.2), and the opacity
encountered by the escaping radiation (Sec.~3.3). In
Sec.~\ref{sec:populations}, we perform population synthesis calculations
aimed at testing the likelihood of our inferred binary parameters. We
summarize and conclude in Sec.~\ref{sec:discussion}.

\section{Delayed emission from the transients AT2018cow, AT2020mrf,
AT2022tsd}\label{sec:at2022tsd}

In the following we summarize the main observations of the three FBOTs
which are known to display late-time emission.

\subsection{AT2018cow}

AT2018cow is a relatively nearby FBOT, exploded on 16 June 2018 at a
distance of $\sim 62$~Mpc in a star-forming dwarf galaxy
\citep{Prentice2018}. Initially detected in optical \citep{Smartt+2018},
it reached a peak luminosity of $\sim 4\times 10^{44}$~erg/s
\citep{Margutti2019} and it was detected over a wide electromagnetic
range, from hard X-rays at energies $> 10$~keV, to the radio band. The
latter was found consistent with the interaction of a blast wave of
$v\sim 0.1c$ with a dense environment \citep{Margutti2019}. However, a
search for $\gamma$-ray emission with the Inter-Planetary network ruled
out emission with peak luminosity $>10^{47}$~erg/s, hence further
differentiating this transient from a standard gamma-ray burst.

After its discovery, AT2018cow was continued to be monitored. Its
luminosity was found to have considerably dropped, by about 4 orders of
magnitude down to $2 \sim 10^{39}$~erg/s, after about 200 days from the
initial discovery \citep{Migliori2024}. However, an X-ray source at a
level lower by only  a factor $\lesssim 4 $  was still present after
about 3.7 years \citep{Migliori2024}. Multiband UV photometry acquired
with the Hubble telescope at around the same time at the location of
AT2018cow revealed the presence of a source with luminosity $L_{\rm UV}
\sim 10^{39}$~erg/s \citep{Chen2023}. The UV spectrum was found to be
consistent with the Rayleigh-Jeans tail of a thermal spectrum with
effective temperature $T_{\rm eff}\gtrsim 10^{4.6}$~K.  Considering the
possbility that the optical, UV and X-ray emission come from the same
origin, \citet{Margutti2019} fitted it with a \citet{Shakura1973}
multibody disk with inner temperature of $T_{\rm in} \sim {10^{5.91}}$~K
and outer temperature $T_{\rm out} \sim {10^{4.45}}$~K.

While the observed late emission could be interpreted within the context
of an accretion disk around an intermediate mass BH
\citep{Migliori2024}, in the following we will entertain the possibility
that it is the result of accretion onto a companion BH, motivated by the
observation that the late-time emission appears of a different nature
than the one of the main LFBOT \citep{Migliori2024}.

\subsection{AT2020mrf}

The FBOT AT2020mrf was discovered on 12 June 2020 by ZTF. Its X-ray,
optical and radio properties were reported by \citet{Yao2022}, together
with the analysis of the host galaxy. Likewise for the AT2018cow, the
host was found to be a dwarf galaxy; its mass was estimated to be
$M_{*}\sim 10^8 M_\odot$, and its specific star formation rate  $\sim
10^{-1}$~yr$^{-1}$.  The metallicity of the galaxy was estimated between
$10^{-0.70}-10^{-0.13}Z_\odot$. The optical spectrum of the transient
was fitted by a blackbody with temperature $T\sim 2\times 10^4$~K and
emitting radius $R\sim 8\times 10^{14}$~cm. The X-ray luminosity, on the
order of $2\times 10^{43}$~erg/s at $\sim 36$~days, is comparable to
that of cosmological GRBs. However, the $\gamma$-ray upper limit of
$\sim 10^{49}$~erg/s  from Konus-Wind is considerably lower than that of
a standard GRB seen on-axis. The bright radio luminosity, at a level of
$\sim 10^{39}$~erg/s, could be explained as synchrotron emission from
the interaction between a blastwave with velocity $v\sim 0.07-0.08~c$
and a dense medium, similarly to AT2018cow.

The source At2020mrf was again detected in X-rays by {\em Chandra} at
328 days, with a luminosity level of  $\sim 10^{42}$~erg/s and with 
intraday variability \citep{Yao2022}. This emission was found to be too
bright to be an extension of the radio synchrotron spectrum. The 
spectrum at 328 days was found to be harder, $F_\nu \propto \nu^0$
compared to the spectrum at $\sim 36$~days, $F_\nu \propto \nu^{-0.8}$.

Similarly to the case of AT2018cow, the properties of AT2020mrf require
an extended energy source, for whom the most natural candidates are
either a magnetar or an accreting BH. If the FBOT is produced by the
result of a successful SN-like explosion, then no significant emission
from fallback is expected at about a year (e.g.
\citealt{Zhang2008,Perna2014}), while a failed explosion would require a
weakly bound red giant progenitor \citep{Yao2022}. In the case of a
remnant magnetar, X-rays may be generated at the termination shock of a
pulsar wind nebula \citep{Lyutikov2022}. In the following we will show
that the late emission of AT2020mrf can be naturally explained within
our scenario.

\subsection{AT2022tsd}

AT2022tsd, also known as the Tasmanian Devil, was discovered on 7
September 2022 by the Zwicky Transient Facility, and was promptly
followed with an array of telescopes, revealing the source across a
broad electromagnetic spectrum from the X-rays to the millimeter 
\citep{Ho2023}. It was localized at about 6~kpc from the center of a
dwarf galaxy.

The presence of a long-lived energy source is particularly evident for
this transient. Starting at about 26~days after the initial transient
discovery, optical photometry revealed flaring activity, persisting for
about 100 days. A total of 14 flares were recorded, with
minute-timescale duration, and occasional luminosity that was nearly as
bright as that of the original transient, $\sim 10^{44}$~erg~s$^{-1}$.
Additionally, in between flares, flux variations exceeding an order of
magnitude on timescales shorter than 20~s were also observed.

The flare analysis performed by \citet{Ho2023} revealed a radius of the
emitting region which is much smaller than the blackbody radius of the
early emission from the LFBOT. The short timescales of the flares,
combined with their large energetics, imply optically thin emission from
an at least mildly relativistic outflow with velocity $v/c\gtrsim 0.6$.
Therefore, they conclude that there must be a long-lived engine
associated with the initial FLBOT transient. As will be shown in the
following, the model proposed here provides a compelling explanation for
the origin of these observed flares.

\section{Late emission from clumpy accretion onto a BH companion}
\label{sec:model}

\subsection{Mass inflow}

\begin{figure*} \includegraphics[width=\textwidth]{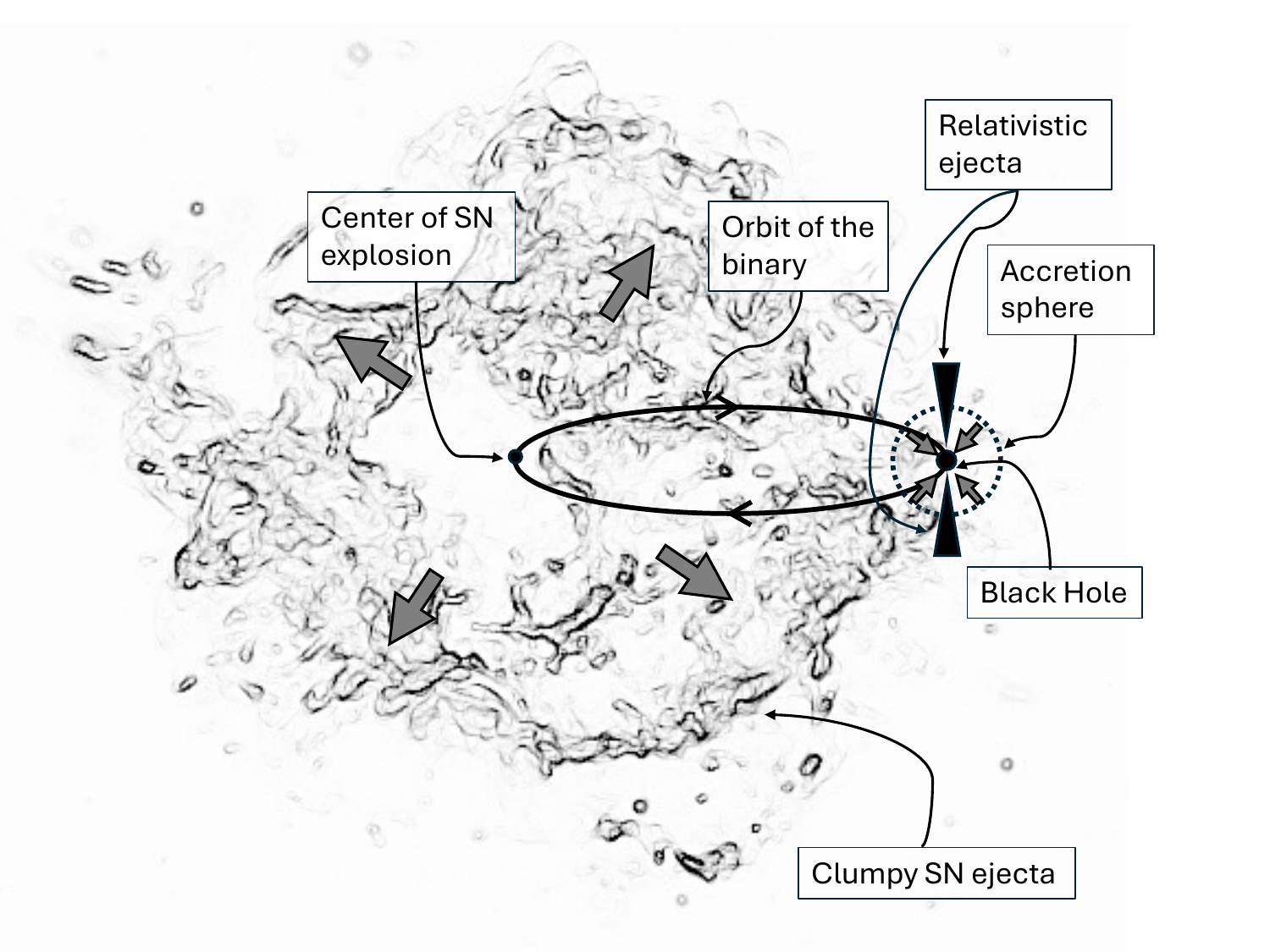}
\caption{Cartoon of the post SN explosion binary system, as envisioned
in this model. The SN ejecta have expanded to reach the orbit of the
companion BH, which is accreting mass via the Hoyle-Littleton mechanism.
Ejecta clumps are shredded into an accretion disk and the accreting BH
is ejecting a collimated relativistic outflow that is eventually
responsible for the observed non-thermal electromagnetic emission.}
\label{fig:cartoon} \end{figure*}

Let us consider a binary system with separation $d$. The system contains
an evolved star that has recently exploded as a core-collapse SN, and a
companion BH.  The exploded star has ejected a mass $M_{\rm{ej}}$ at a
velocity $v_{\rm{ej}}$\footnote{Most likely the ejecta have a spread of
velocities. See the discussion for an evaluation of the impact of such
occurrence.} (see a graphical representation of this scenario in
Figure~\ref{fig:cartoon}, and also \citealt{Kimura2017a,Kimura2017b} for
a similar scenario). For the case of uniform ejecta we follow closely
the derivation in \cite{Kimura2017b}. We then extend the formalism by
considering clumping of the ejecta. The ejecta reach the BH after a time

\begin{equation} \Delta t_{\rm{SN-trans}}=\frac{d}{v_{\rm{ej}}} = 17 \,
\left( \frac{d}{10~\rm{AU}}\right) \left(
\frac{v_{\rm{ej}}}{10^{8}\frac{\rm{cm}}{\rm{s}}}\right)^{-1}\;\;\;
\mathrm{d}\,. \label{eq:tobs} \end{equation}

Since the ejecta velocity is likely to greatly exceed their sound speed,
the cross section for accretion onto the companion BH is given by the
Hoyle-Littleton radius (e.g., \citealt{Edgar2004}):

\begin{equation} r_{\rm{acc}}=\frac{2GM_{\rm{BH}}}{v_{\rm{ej}}^2} =
4\times10^{11}  \left(\frac{M_{\rm{BH}}}{15M_\odot}\right) \left(
\frac{v_{\rm{ej}}}{10^{8}\frac{\rm{cm}}{\rm{s}}}\right)^{-2} \;
\mathrm{cm}\,, \label{eq:racc} \end{equation}

\noindent where $M_{\rm{BH}}$ is the mass of the BH. The total accreted
mass after the entire remnant has crossed the BH orbit is therefore:

\begin{table}
\centering 
\begin{tabular}{lcr} \hline 
Physical parameter & symbol & fiducial value  \\ \hline 
orbital diameter & $d$& 10 AU \\ 
velocity of the SN ejecta & $v_{\rm{ej}}$ & $10^8$ cm s$^{-1}$ \\
mass of the companion BH & $M_{\rm{BH}}$ & $15\,M_\odot$ \\ 
mass of the SN ejecta & $M_{\rm{ej}}$ & $1\,M_\odot$ \\ 
thickness of the ejecta & $\Delta$ & $d$ \\ 
volume filling factor of ejecta clumps & $\zeta$ & $10^{-3}$ \\ 
clump radius relative to ejecta & $\xi$ & $10^{-3}$  \\ 
BH ejection efficiency & $\eta$ & 0.1 \\ 
Impact parameter of individual clump & $r_0$ & $4\times10^{11}$ cm\\ 
\hline 
\end{tabular} 
\caption{Physical parameters of the model and their fiducial values used for
order of magnitude estimates.} 
\label{tab:params} 
\end{table}

\begin{equation} M_{\rm{acc}} = M_{\rm{ej}}\frac{r_{\rm{acc}}^2}{4\,d^2}
= M_{\rm{ej}} \frac{(GM_{\rm{BH}})^2}{d^2 v_{\rm{ej}}^4} =
1.8\times10^{-6} \, \left(\frac{M_{\rm{ej}}}{M_\odot}\right)
\left(\frac{M_{\rm{BH}}}{15M_\odot}\right)^2 \left(
\frac{d}{10~\rm{AU}}\right)^{-2} \left(
\frac{v_{\rm{ej}}}{10^{8}\frac{\rm{cm}}{\rm{s}}}\right)^{-4}
\;\;M_\odot\,. \end{equation} \noindent

If we consider a remnant with radial thickness $\Delta$, an estimate of
the average accretion rate yields

\begin{equation} 
\bar{\dot{M}}_{\rm{acc}} =
\frac{M_{\rm{acc}}v_{\rm{ej}}}{\Delta} = M_{\rm{ej}} \frac{(G
M_{\rm{BH}})^2}{d^2 \Delta  \,v_{\rm{ej}}^3} = 10^{-12} \frac{d}{\Delta}\;
\left(\frac{M_{\rm{ej}}}{M_\odot}\right)\left(\frac{M_{\rm{BH}}}{15M_\odot}\right)^2 
\left( \frac{d}{10~\rm{AU}}\right)^{-3} \left(
\frac{v_{\rm{ej}}}{10^{8}\frac{\rm{cm}}{\rm{s}}}\right)^{-3}
\;\;\frac{M_\odot}{\rm s}\,. \label{eq:mdot_ave} 
\end{equation}

We note that this rate is highly super-Eddington, the Eddington
accretion rate onto a $15~M_\odot$ black hole being
$\sim10^{-15}/\eta~(M_\odot$/s), where $\eta$ represents an efficiency
for the conversion of accreted rest-mass $\bar{\dot{M}}_{\rm{acc}}c^2$
into radiated luminosity. The inflow is also likely to form an accretion
disk because of the orbital angular momentum
\citep{deVal-Borro2009,HuarteEspinosa2013}. Accretion disks onto BHs
with super-Eddington accretion rates are known, from both theoretical
investigations (e.g. \citealt{Blandford1999}), numerical simulations
(e.g. \citealt{Takeuchi2013,Jiang2014}), and observations (e.g.
\citealt{Fabrika2004}) to feature strong outflows. If we make this
assumption here, and indicate with $\Omega$ the outflow collimation
angle, we obtain an average observed bolometric luminosity: 

\begin{equation} \bar{L}_{\rm{iso}} = \eta
\frac{4\pi}{\Omega}\bar{\dot{M}}_{\rm{acc}} c^2 \simeq 2\times10^{42}
\eta \frac{4\pi}{\Omega} \frac{d}{\Delta} \;
\left(\frac{M_{\rm{ej}}}{M_\odot}\right)
\left(\frac{M_{\rm{BH}}}{15M_\odot}\right)^2 \left(
\frac{d}{10~\rm{AU}}\right)^{-3} \left(
\frac{v_{\rm{ej}}}{10^{8}\frac{\rm{cm}}{\rm{s}}}\right)^{-3} \;\;
\frac{\mathrm{erg}}{\mathrm{s}}\,, \label{eq:Lave} \end{equation}

which is a considerable energy output, comparable to the supernova
luminosity at peak.

As discussed above, a more accurate model would be one in which the SN
ejecta are clumped, characterized by a clump volume filling factor
$\zeta$ and by a typical clump radius $r_c=\xi d$. In this case, the
emitted power is more accurately calculated as the mass of a single
clump over the time it takes for the clump to be accreted via a disk by
the central black hole, modified with the efficiency and geometric
beaming factors as above. We now explore this scenario in more detail.

The total number of clumps in the remnant is given by

\begin{equation} N_c = \frac{3d^2\Delta}{r_c^3}\zeta = 3
\frac{\zeta}{\xi^3}\frac{\Delta}{d}=  3\times10^6 \frac{\Delta}{d} \;
\left(\frac{\zeta}{10^{-3}}\right) \left(
\frac{\xi}{10^{-3}}\right)^{-3}\,, \end{equation}

where the fiducial values for $\zeta$ and $\xi$ are taken from both
observational (e.g. \citealt{Fransson1989}) and theoretical (e.g.
\citealt{Dessart2018}) studies. Of all such clumps, the number of those
that fall within the BH accretion radius reads:

\begin{equation} N_{\rm{acc}} = \frac{\pi r_{\rm{acc}}^2}{4\pi d^2} N_c
= 3 \left(\frac{GM_{\rm{BH}}}{d \,v_{\rm{ej}}^2}\right)^2 
\frac{\zeta}{\xi^3} \, \frac{\Delta}{d} =  5 \, \frac{\Delta}{d}
\left(\frac{M_{\rm{BH}}}{15 M_\odot}\right)^2
\left(\frac{d}{10~\rm{AU}}\right)^{-2} \left(\frac{v_{\rm{ej}}}{10^{8}
\frac{\rm{cm}}{\rm{s}}}\right)^{-4} \left(\frac{\zeta}{10^{-3}}\right)
\left( \frac{\xi}{10^{-3}}\right)^{-3}\,. \label{eq:nacc} \end{equation}

Clump accretion can take place under two distinct regimes, depending on
the system properties. For small orbital separation, it is likely that
the clumps are smaller than the accretion radius in Eq.~\ref{eq:racc},
while at larger separations partial accretion of big clumps would be
possible. The orbital separation at which large clump accretion becomes
relevant is

\begin{equation} d> 2\frac{GM_{\rm{BH}}}{\xi v_{\rm{ej}}^2} = 26
\left(\frac{M_{\rm BH}}{15\Msol}\right)
\left(\frac{\xi}{10^{-3}}\right)^{-1} \left(\frac{v_{\rm
ej}}{10^8\frac{\rm{cm}}{\rm{s}}}\right)^{-2} \;\;\mathrm{AU}\,.
\label{eq:largeClumps} \end{equation} We note that if
Eq.~\ref{eq:largeClumps} is satisfied, then the number of accreted
clumps falls below unity (see Eq.~\ref{eq:nacc}) and we therefore
concentrate on the small clumps regime. At large radii, the accretion is
likely to be dominated by the intra-clump ejecta, following
Eqs.~\ref{eq:mdot_ave} and~\ref{eq:Lave}.

Let us consider a small clump approaching the accretion region with an
impact parameter $r_0$ (so $r_0\lesssim r_{\rm acc}$) around the BH and
with velocity $v_{\rm ej}$. When a clump enters the Hoyle-Littleton
sphere of the BH, the clump  falls towards the BH on the free-fall
timescale:

\begin{equation} t_{\rm ff} \approx \sqrt{\frac{r_{\rm acc}^{3}}{GM_{\rm
BH}}} = 2\sqrt{2}\,\frac{GM_{\rm{BH}}}{v_{\rm{ej}}^3} \simeq
1.5\left(\frac{M_{\rm
BH}}{15\Msol}\right)\left(\frac{v_{\rm{ej}}}{10^{8}
\frac{\rm{cm}}{\rm{s}}}\right)^{-3}{\rm hour}. \end{equation}

If the angular momentum loss is negligible in its decaying orbit, the
clump would circularize (once  shredded as a result of tidal forces and
pressure gradients during the infall towards the BH)  at a radius
$r_{\rm circ}$ at which its specific angular momentum equals the one of
a Keplerian orbit, or $r_0\,v_{\rm ej}\sim r_{\rm circ} \sqrt{GM_{\rm
BH}/r_{\rm circ}}$.  This relation yields:

\begin{equation} r_{\rm circ} = \frac{r_0^2v^2_{\rm ej}} {GM_{\rm BH}} =
2\times 10^{11}\left(\frac{M_{\rm{BH}}}{15
M_\odot}\right)^{-1}\left(\frac{r_0}{4\times}10^{11}{\rm cm}\right)^2
\left(\frac{v_{\rm{ej}}}{10^{8} \frac{\rm{cm}}{\rm{s}}}\right)^2{\rm
cm}\,. \label{eq:rcirc} \end{equation}

Note that in realistic situations, infalling streams are bent around the
BH and collide with each other, resulting in deflecting some of the
streams towards the BH. For such cases, the circularization radius would
be smaller than our crude estimate of $r_{\rm circ}$. In this sense,
$r_{\rm circ}$ in Equation~\ref{eq:rcirc} should be considered as an
upper limit.  Lower values would yield shorter timescales for accretion
and hence higher peak luminosities (see below).

Upon circularization, the material will accrete onto the BH on the
viscous timescale, which reads (for a puffy disk, as typical of
hyperaccretion):

\begin{equation} t_{\rm visc} \simeq \frac{1}{\alpha} \left(\frac{r_{\rm
circ}^3}{G M_{\rm BH}}\right)^{1/2} = \frac{1}{\alpha} \frac{\left(r_0
\, v_{\rm{ej}}\right)^3}{\left(G\, M_{\rm{BH}}\right)^2} \simeq
1.5\times10^{5} \left(\frac{\alpha}{0.1}\right)^{-1} \left(\frac{M_{\rm
BH}}{15M_\odot}\right)^{-2}\left(\frac{r_0}{4\times10^{11}{\rm
cm}}\right)^{3} \left(\frac{v_{\rm{ej}}}{10^{8}
\frac{\rm{cm}}{\rm{s}}}\right)^{3}{\rm s}\,, \label{eq:tvisc}
\end{equation}

where $\alpha$ is the viscosity parameter \citep{Shakura1973}. The clump
accretion rate may thus be estimated as

\begin{eqnarray} \label{eq:mdot_vis} \dot{M}_{\rm c} &\simeq&
\frac{M_{\rm c}}{t_{\rm visc}} = \frac{\alpha}{3} \frac{d}{\Delta}
\frac{\xi^3}{\zeta} G^2 \frac{M_{\rm{ej}}\,M_{\rm{BH}}^2}
{r_0^3\,v_{\rm{ej}}^3} \nonumber \\ &\simeq& 2\times
10^{-12}\frac{d}{\Delta} \left(\frac{\alpha}{0.1}\right)
\left(\frac{\xi}{10^{-3}}\right)^3
\left(\frac{\zeta}{10^{-3}}\right)^{-1}
\left(\frac{M_{\rm{ej}}}{1\,M_\odot}\right) \left(\frac{M_{\rm
BH}}{15M_\odot}\right)^{2}\left(\frac{r_0}{4\times10^{11}{\rm
cm}}\right)^{-3} \left(\frac{v_{\rm{ej}}}{10^{8}
\frac{\rm{cm}}{\rm{s}}}\right)^{-3}\;\;\frac{M_\odot}{\rm s}\,,
\label{eq:mdotClump} \end{eqnarray} where we have estimated the clump
mass as $M_c=M_{\rm{ej}}/N_c$, which is valid if the clumps carry the
majority of the ejecta mass. Note that the viscous time and,
consequently, the accretion rate have a strong dependence on the impact
parameter $r_{\rm 0}$ and ejecta velocity $v_{\rm ej}$. For a given
ejecta velocity, the accretion rate of the clumps with small impact
parameter $r_0$ is going to be much higher than the one of the --- far
more numerous --- clumps that are accreted with an impact parameter
comparable to the BH accretion radius $r_{\rm{acc}}$. The accretion rate
history of the BH is therefore characterized by  a low-level steady
state from the tenuous intra-cluster ejecta material, numerous
long-lasting low-level events from clumps accreted with $r_0\sim
r_{\rm{acc}}$, and few strong events when a low impact parameter clump
is accreted. 
%(see also \citealt{Perna2006} for a similar flare model produced from
%clumpy accretion in $\gamma$-ray bursts accretion disks). 

To evaluate the maximum expected accretion rate, we consider that the
clump with the smallest impact parameter has $r_0\sim r_c$. In addition,
we calculate the number of clumps that are expected to accrete with an
impact parameter compared to their size to be:

\begin{equation} N_{\rm{c},r_0=r_{\rm c}} = N_{\rm c} \frac{\pi r_{\rm
c}^2}{4\pi d^2}= \frac34 \frac\zeta\xi \frac\Delta{d}\,, \end{equation}

which is of order unity for the fiducial model parameters described in
Table~\ref{tab:params}. The maximum expected accretion rate
(Eq~\ref{eq:mdotClump} at $r_{0}=r_{\rm c}$) is therefore:

\begin{equation} \dot{M}_{\max} = \frac{\alpha}{3\zeta} \frac{d}\Delta
G^2 \frac{M_{\rm{ej}}\, M_{\rm{BH}}^2}{d^3\,v_{\rm{ej}}^3} =
4\times10^{-11} \frac{d}{\Delta} \left(\frac{\alpha}{0.1}\right)
\left(\frac{\zeta}{10^{-3}}\right)^{-1}
\left(\frac{M_{\rm{ej}}}{1\,M_\odot}\right) \left(\frac{M_{\rm
BH}}{15M_\odot}\right)^{2}\left(\frac{d}{10\,{\rm{AU}}}\right)^{-3}
\left(\frac{v_{\rm{ej}}}{10^{8}
\frac{\rm{cm}}{\rm{s}}}\right)^{-3}\;\;\frac{M_\odot}{\rm s}\,.
\label{eq:mdotMax} \end{equation}

As mentioned above, given the highly super-Eddington accretion rates,
and the suitable conditions for circularization of the clump ejecta, it
is presumable that a beamed and  moderately relativistic outflow may be
launched (e.g.,
\citealt{Blandford1977,Tchekhovskoy2011,Qian2018,Parfrey2019}). Let its
opening angle be $\Omega$. We can then estimate the isotropic equivalent
kinetic energy that is ejected from the BH as a result of the accretion
of a clump, which can be potentially radiated as an electromagnetic
transient:

\begin{equation} {L}_{\rm{c,iso}} = \eta
\frac{4\pi}{\Omega}\dot{M}_{\rm{c}} c^2 \simeq 3\times10^{42} \, \eta
\frac{4\pi}{\Omega} \frac{d}{\Delta} \left(\frac{\alpha}{0.1}\right)
\left(\frac{\xi}{10^{-3}}\right)^3
\left(\frac{\zeta}{10^{-3}}\right)^{-1}
\left(\frac{M_{\rm{ej}}}{1\,M_\odot}\right) \left(\frac{M_{\rm
BH}}{15M_\odot}\right)^{2}\left(\frac{r_0}{4\times10^{11}{\rm
cm}}\right)^{-3} \left(\frac{v_{\rm{ej}}}{10^{8}
\frac{\rm{cm}}{\rm{s}}}\right)^{-3} \;\;\;
\frac{\mathrm{erg}}{\mathrm{s}}\,. \end{equation}

The maximum luminosity is correspondingly given by

\begin{equation} {L}_{\rm{\max,iso}} = \eta
\frac{4\pi}{\Omega}\dot{M}_{\max} c^2 \simeq 7\times 10^{43} \, \eta
\frac{4\pi}{\Omega} \frac{d}{\Delta} \left(\frac{\alpha}{0.1}\right)
\left(\frac{\zeta}{10^{-3}}\right)^{-1}
\left(\frac{M_{\rm{ej}}}{1\,M_\odot}\right) \left(\frac{M_{\rm
BH}}{15M_\odot}\right)^{2}\left(\frac{d}{10\,{\rm{AU}}}\right)^{-3}
\left(\frac{v_{\rm{ej}}}{10^{8} \frac{\rm{cm}}{\rm{s}}}\right)^{-3}
\;\;\; \frac{\mathrm{erg}}{\mathrm{s}}\,. \label{eq:Lmax} \end{equation}

If we now substitute the clump radius in Eq.~\ref{eq:tvisc}, we
obtain the minimum variability timescale, which is associated with the
brightest flares, as: 
\begin{equation} 
t_{\rm{var,\,min}}=\frac{1}{\alpha} \frac{(\xi
\,d \,v_{\rm{ej}})^3}{(GM_{\rm{BH}})^2}=2
\left(\frac{\alpha}{0.1}\right)^{-1} \left(\frac{\xi}{10^{-3}}\right)^3
\left(\frac{d}{10\,{\rm{AU}}}\right)^{3} \left(\frac{v_{\rm{ej}}}{10^{8}
\frac{\rm{cm}}{\rm{s}}}\right)^{3} \left(\frac{M_{\rm
BH}}{15M_\odot}\right)^{-2} \;\;\; \mathrm{h}\,. \label{eq:tvarmin}
\end{equation}

We notice that the maximum luminosity in Eq.~\ref{eq:Lmax}
scales with the binary properties analogously to the average luminosity
in Eq.~\ref{eq:Lave}, so that we can write:

\begin{equation} {L}_{\rm{\max,iso}} = \frac{\alpha}{3\zeta}
\frac{d}\Delta \bar{L}_{\rm{iso}} \simeq 33
\left(\frac{\alpha}{0.1}\right) \left(\frac{\zeta}{10^{-3}}\right)^{-1}
\frac{d}\Delta \, \bar{L}_{\rm{iso}}\,. \label{Lratio} \end{equation}

\begin{figure}
\centerline{\includegraphics[width=0.6\textwidth]{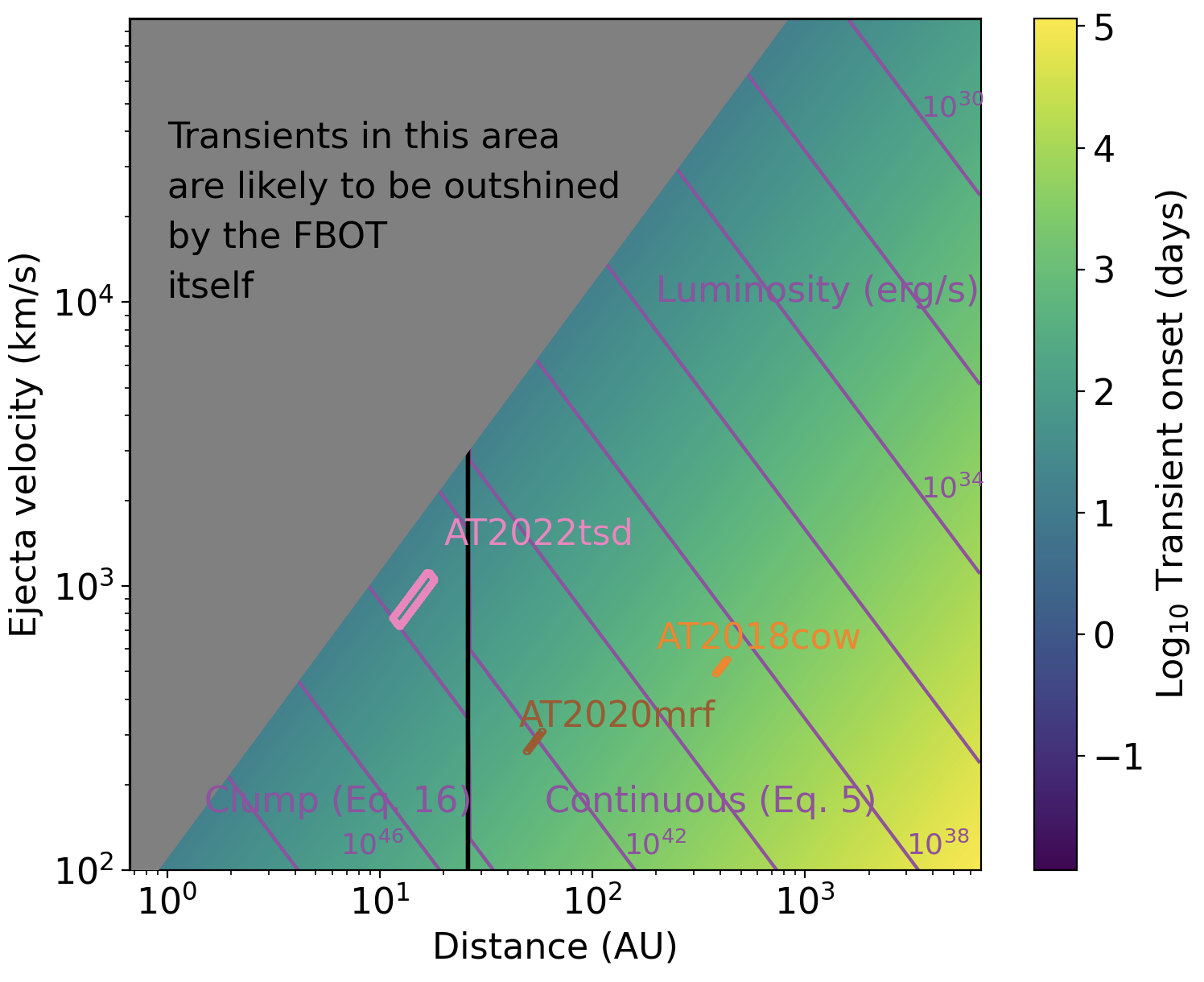}}
\caption{Delay and luminosity of the BH-powered transients from
accretion induced by a supernova in the binary system as a function of
the binary separation and the ejecta velocity. Time delay is shown in
pseudocolors (see colorbar), while average luminosity from
Eqs.~\ref{eq:Lave} and~\ref{eq:Lmax} is shown with contour levels. A
vertical black line separates the region in which accretion from
individual clumps dominates (left, Eq.~\ref{eq:Lmax}) and the region in
which accretion is dominated by the intraclump medium. The location in
the parameter space of the late emission from the three considered FBOTs
is also shown. Except from binary separation $d$ and ejecta velocity
$v_{\rm{ej}}$, all parameters are set to their fiducial values shown in
Table~\ref{tab:params}.}\label{fig:theThree} \end{figure}

Fig.~\ref{fig:theThree} shows,  for the fiducial parameters in Table~1,
how the delay and the  luminosity of the accreting primary BH vary as a
function of the binary BH orbital distance and the velocity of the SN
ejecta. Note the degeneracy between these two variables both in
reproducing a luminosity value as well as a time delay. However, the
different functional form of the correlation allows one to identify
specific regions in the $\{d,v_{\rm ej}\}$ parameter space when
observations provide constraints on both luminosity and time delay.

The vertical line in the figure separates the region to the left, where
individual clumps dominate, and hence flaring emission is more likely
expected with the enhanced luminosity of Eq.~\ref{eq:Lmax}, from the
region to the right, where the emission is more likely to be continuous
and hence the luminosity is expected to be closer to the average value
of Eq.~\ref{eq:Lave}. The location of the line is set by
Eq.~\ref{eq:largeClumps}.

Last, the gray area in the upper left corner indicates a region in which
transients from the accreting companion BH would likely be outshined by
(or at least confused with) the primary emission from the FBOT itself.
This has been taken corresponding to a time delay between the FBOT and
the later emission of $\leq 2$~weeks.

Examining more in detail Fig.~\ref{fig:theThree} for each of the
sources, we see that, for the case of AT2018cow, a luminosity on the
order of $\sim 10^{39}$~erg/s with a time delay of 3.7 years can be
produced with a BH companion at a distance of about 400~AU, accreting SN
ejecta moving at a speed of $\sim 500$~km/s. For this parameter
set, accretion is dominated by the intra-clump medium (cfr.
Eq.~\ref{eq:largeClumps}), and pronounced variability is not expected.
The case of AT2018cow is, however, unique among the FLBOTs discussed
here since excess X-ray emission has been detected at multiple times
with no indication of fast variability. If, for example, the component
seen at 220~days is interpreted as the beginning of the accretion phase
onto the companion BH, then the point corresponding to AT2018cow in
Fig.~\ref{fig:theThree} would move along a track parallel to the blue
lines, to an orbital distance of about 100~AU and an ejecta velocity of
about $2\times 10^3$~km/s. Either way, the region in the parameter space
would correspond to one in which the emission is not expected to display
flaring due to clumps (consistent with observations). We also note that
our inferred ejecta speeds, when combined with the blastwave velocity of
$\sim 0.1 c$ required to explain the early radio emission from the FBOT
\citep{Margutti2019}, would indicate an asymmetric SN explosion and/or
that the mildly relativistic component whose interaction with the dense
medium produces the FBOT radio emission is powered by the outflow/jet
which emerges perpendicularly to the orbital plane, where the slower SN
ejecta move.

For the case of AT2020mrf, the detected emission at 326 days with a
luminosity of $\sim 10^{42}$~erg/s places the accreting BH companion an
an orbital distance of about 50~AU and ejecta speed of about 400~km/s.
However, we note that the extended X-ray component could have started
earlier, any time between 36-days - 336 days, when there were no
observations. This would move the point in Fig.~\ref{fig:theThree} along
the blue line corresponding to $10^{42}$~erg/s, reaching at $t=75$~d the
border with the region where variable, late-emission is expected.

Last, as shown in Fig.~\ref{fig:theThree} for the source AT2022tsd, for
our fiducial choices of BH companion mass and SN ejecta properties, the
flare luminosity of $\sim 10^{44}$~erg~s$^{-1}$, and the $\sim 1$~month
delay of the late emission can be reproduced for a BH binary with an
orbital separation of $\sim15$~AU and ejecta velocity $\sim 10^3$~km/s.

Taking this parameter set at face value, Eq.~\ref{eq:tvarmin} gives a
variability timescale of several hours, in excess of the observed
timescale of tens of minutes. This tension between the model and the
observations can be easily ameliorated by slightly modifying the adopted
fiducial model parameters, such as increasing the accreting BH mass or
decreasing the size of the clumps. For example, reducing by a factor 3
the clumping factor to $\xi=3\times10^{-4}$ (with all the other
parameters remaining the same), would yield a variability timescale of
$\sim15$~minutes. In addition, instabilities in the accretion and/or
substructure in the clumps geometry would result in variable luminosity
on shorter timescales than the order of magnitude estimate provided by
Eq.~\ref{eq:tvarmin}.

We finally note that, in all the cases, the inferred ejecta
velocities are on the lower end for core-collapse supernovae. However,
in our model the companion BH receives the equatorial ejecta from the
exploding star, while the supernova itself is likely observed from the
polar direction. If FLBOTs are engine driven, as suggested by numerous
models (e.g., \citealt{Margutti2019, Perley2019}), a polar dependence of
the supernova ejecta is to be expected. In addition, the emission from
the companion BH is likely to be dominated by slower ejecta
(Eq.~\ref{eq:Lmax}), which carry more mass than the ejecta that are in
the leading edge of a homologous expansion (see, e.g.,
\citealt{Chevalier1989,Lazzati2012}).

\begin{figure} \centering
\includegraphics[width=0.32\textwidth]{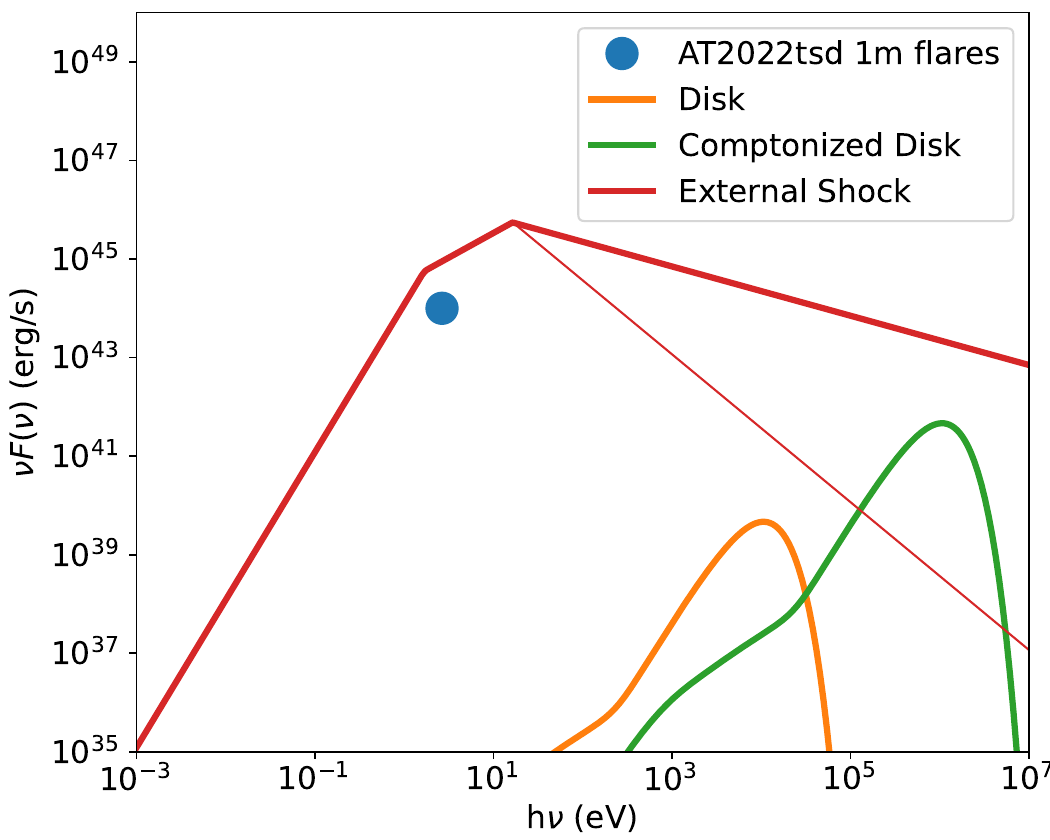} 
\includegraphics[width=0.32\textwidth]{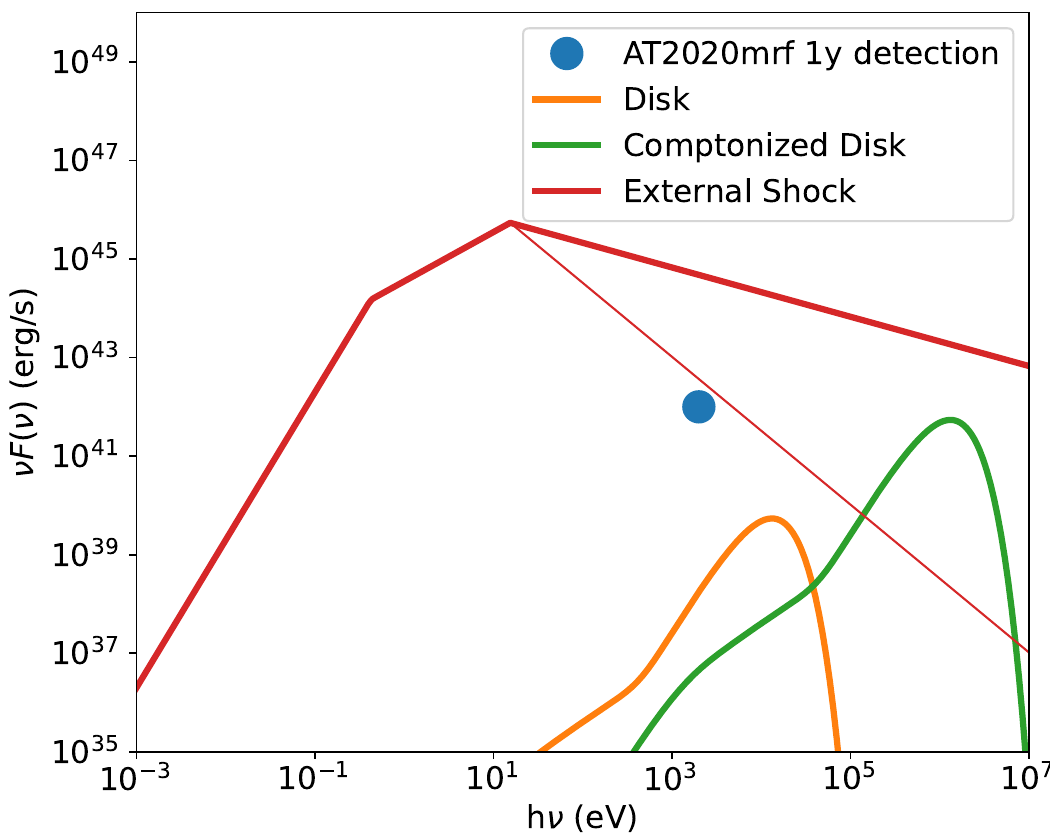}   
\includegraphics[width=0.32\textwidth]{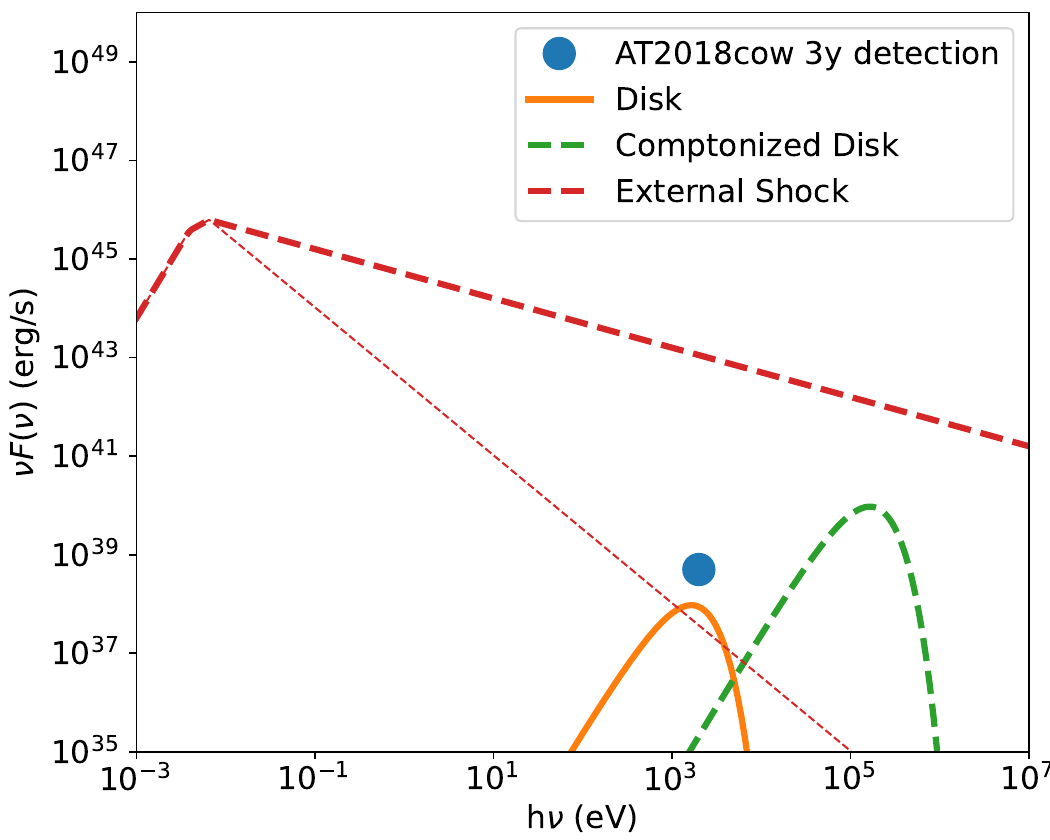} \caption{Various
components of the emission spectrum of a BH accreting from SN ejecta: an
accretion disk, Comptonized disk emission, and radiation from an
external shock formed from the interaction of an accretion-driven
outflow with the dense SN medium. Dashed lines are used to emphasize
that an accretion-driven outflow might not form when the accretion is
sub-Eddington. A thin line is used, instead, for the high-frequency
slope of the external shock component in case of a steeper electron
distribution. For each source, the model parameters are varied to
produce the best match to the late-time observations (see text).}
\label{fig:emission} \end{figure}

\subsection{Electromagnetic emission} \label{sec:emission}

Let us now consider the electromagnetic signal that is produced by the
accreting companion BH. Since the accreting material forms a disk, we
first consider the emission coming from the disk itself. Accretion rates
are in most cases highly super-Eddington, therefore we compute the disk
emission using the effective temperature profile derived for
hyper-accreting conditions \citep{Kawaguchi2003}:

\begin{equation} T_{\rm eff}(r) = 1.8\times 10^8 \left(\frac{M_{\rm
BH}}{15M_\odot}\right)^{-1/4} \left(\frac{\dot{M}_{\rm acc}c^2}{10^4
L_{\rm Edd}}\right)^{1/4} \left(\frac{R}{R_{\rm Sch}}\right)^{-3/4}
\left(1- \sqrt{\frac{3R_{\rm Sch}}{R}}\right)^{1/4}~{\rm K}\,,
\label{eq:Teff} \end{equation}

where $R_{\rm Sch}$ is the Schwarzschild radius, and where the inner and
outer disk radii are assumed to be $3 R_{\rm Sch}$ and the
circularization radius of Eq.~\ref{eq:rcirc}, respectively. The emission
spectrum is computed by integrating over radius, assuming local black
body properties. The overall spectrum is then normalized to the
luminosity values reported in \cite{Kawaguchi2003} (see their Figure
12), who properly take into account the effect of opacity and
relativistic redshift in their calculations. This component is shown in
Figure~\ref{fig:emission} with a solid orange line. We notice, in
addition, that the disk would likely eject also a subrelativistic wind,
especially if the mass inflow rate is super-Eddington (e.g.,
\citealt{Blandford1999}). If this wind is optically thick, the disk
radiation would be reprocessed in the wind. As a consequence, it would
be converted to shorter wavelengths and the total energetics decreased
due to adiabatic losses \citep{Piro2020}.

In addition to the disk emission, a collimated relativistic outflow is
expected to be generated by the accreting BH, as discussed in Sec.3.1.
Here we consider an internally cold outflow with a mild to moderate
Lorentz factor $\Gamma\sim10$ beamed into a solid angle $\Omega$. This
outflow would produce EM emission in two ways. First, by bulk
Comptonization of the disk photons, also known as Compton Drag (CD), and
secondly by synchrotron emission, upon impacting the dense supernova
remnant material.

Let us first consider the CD emission. The Comptonized spectrum can be
easily calculated by boosting each photon's frequency by a factor
$\Gamma^2$ \citep{Rybicki1986,Lazzati2000}. In a steady-state scenario,
like the one envisaged here, the bolometric luminosity is boosted by a
factor $L_{\rm{bol, CD}}=\min[\tau_T,1] L_{\rm{bol,seed}}\Gamma^2$,
where $\tau_T$ is the Thomson optical depth of the relativistic outflow
and $L_{\rm{bol,seed}}$ is the luminosity of the seed photons, in our
case the disk emission. The resulting spectra are shown in
Figure~\ref{fig:emission} with a green line, under the assumption
$\tau_T\ge1$. While this assumption may not always be realized, we
considered carrying out a full integration to be beyond the scope of
this work, given that the CD emission is never the one matching the
observations in the three considered transients. We also assumed
$\Gamma$ to be constant, while the outflow might be accelerating at the
smallest distances. The green lines in Figure~\ref{fig:emission} should
therefore be considered as an upper limit for the CD emission.

Finally, let us consider the emission from the jet itself, as it strikes
the external medium driving relativistic shocks both into the external
medium and into the jet itself (reverse shock). This emission component
is analogous to what is observed in gamma-ray bursts (GRBs). The outflow
propagates in the high density external medium of the SN ejecta,

\begin{equation} n_{\rm{SNR}} = \frac{3 \zeta M_{\rm{ej}}}{4\pi\,m_p
d^3} = 10^{14} \, \frac{M_{\rm{ej}}}{M_\odot} \frac{\zeta}{10^{-3}}
\left(\frac{d}{1 \,{\rm AU}}\right)^{-3} \;\;\; {\rm cm}^{-3}
\end{equation}

and therefore produces electromagnetic emission through its interaction
with the external material, akin to the case of gamma-ray bursts in very
high density media described in \cite{Wang2022} and \cite{Lazzati2022}.
In this regime the emission is predominantly driven by the repetitive
shocks that the outflow shells produce hitting the external medium
\citep{Lazzati2022,Lazzati2023}, rather than being produced as in the
classic scenario in which shells hit each other first (the so-called
internal shocks, \citealt{Rees1994}). We used the semi-analytical code
developed by \cite{Lazzati2022,Lazzati2023} to compute the synchrotron
emission. The outflow was assumed to be made of 10 shells, with a total
engine activity duration corresponding to the viscosity timescale of
Eq.~\ref{eq:tvisc}. Equipartition parameters $\epsilon_e=0.2$ and
$\epsilon_B=0.01$ were adopted (e.g., \citealt{Panaitescu2002}). To best
mimic the prompt emission of GRBs, the spectrum was assumed to be flat
($F(\nu)\propto\nu^0$) between the self-absorption and peak frequencies.
At higher frequencies, two slopes are shown, to encompass the
variability seen in gamma-ray bursts: $F(\nu)\propto\nu^{-1.5}$ is shown
with a thick line, and $F(\nu)\propto\nu^{-2.5}$ is shown with a thin
line. This component is shown in Figure~\ref{fig:emission} with a red
line. The three panels show the calculated spectra for the tree
considered transients. All parameter values are kept at their fiducial
values, except the orbital separation $d$ and the ejecta velocity
$v_{\rm{ej}}$, which are tuned to the individual transient values
inferred from Figure~\ref{fig:theThree}. In the AT2018cow panel (right)
the outflow emission components are shown with dashed lines since the
accretion rate is comparable to the Eddington value (depending on
$\eta$) and the presence of a collimated outflow is therefore uncertain.
For this specific case, therefore, we concur with the disk emission
interpretation put forward by \cite{Migliori2024}. In our model,
however, the disk surrounds the companion BH and is not the same engine
responsible for the FLBOT explosion. We also notice that a BH more
massive than 15 M$_\odot$ would be required to make the disk emission
consistent with the observed luminosity. Alternatively, the observed
X-ray emission could be interpreted as due to the external shock, if a
jet indeed forms at accretion rates comparable with the Eddington rate.

\subsection{Opacity and emerging radiation} \label{sec:opacity}

Once the EM radiation is produced, it needs to propagate through the SN
ejecta. To compute the column density of material that the transient
radiation needs to propagate through, we adopt the model developed for
SN1987A \citep{Chevalier1989,Chevalier1992}. This model is based on
homologous expansion and is parameterized on the remnant mass and
kinetic energy. It consists of a broken power-law, the break position
propagating outward and smoothly connecting a shallow slope at low radii
($\rho_{\rm{in}}\propto r^{-m}t^{m-3}$) with a steep slope at large
distances ($\rho_{\rm{out}}\propto r^{-n}t^{n-3}$). Following
\cite{Chevalier1992} we set $m=1$ and $n=9$. In our specific calculation
we also set the  velocity of the power-law break in the density profile
to match the evaluated expansion speed for each of the three considered
FBOTs, and we hold the ejecta mass fixed to a solar mass. To convert
column densities to optical depths we use mean opacity values for solar
metallicity gas. For the optical/NIR regime we adopt
$\kappa_{\rm{O/NIR}}=10^{-3}$~g/cm$^{2}$, which is fairly typical for
supernova explosion calculations (e.g.,
\citealt{Ensman1988,Woosley2007}). For the X-ray regime, instead, we
consider a solar metallicity gas in which all the K-shell electrons for
$Z\ge 6$ are bound to their nuclei, which is relevant for gas
temperatures of less than $\sim10^{5}$~K. This gives (e.g.,
\citealt{Lazzati2002}) $\kappa_{\rm{X}}\simeq200$~g/cm$^{2}$.

\begin{figure} \centering
\includegraphics[width=0.6\textwidth]{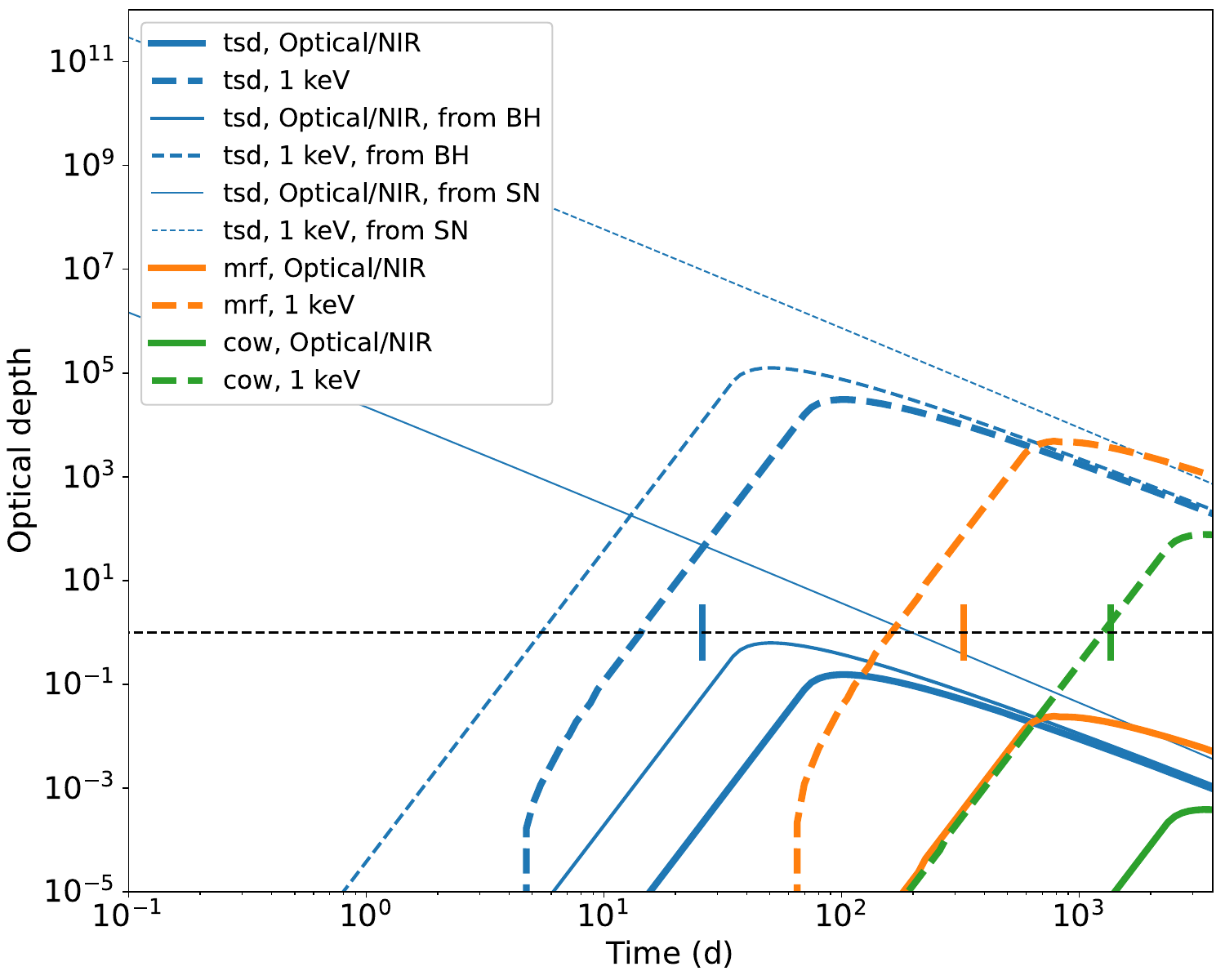} \caption{Opacity
between the transient emission site and the observer at infinity for
different scenarios in both the optical/NIR and X-ray bands (see
Sect.~\protect{\ref{sec:opacity}} for more details).}
\label{fig:optDepths} \end{figure}

The resulting optical depths are shown in Figure~\ref{fig:optDepths},
with colors differentiating the three transients. For each case, the
figure shows several scenarios: solid lines are computed for the
optical/NIR band, while dashed lines refer to the X-ray band. For all
the transients, thick lines refer to the opacity for a transient that is
formed at the location of the external shock of the BH outflow. That is,

\begin{equation} \tau = \kappa \int_{R_{\rm{ES}}}^{\infty} \rho(r')
dr'\,, \label{eq:tau1} \end{equation}

where $r'$ is the distance from the center of the BH and $R_{\rm{ES}}$
is computed via

\begin{equation} M_{\rm{c}}c^2 = \Gamma^2\int_0^{R_{\rm{ES}}}
4\pi\rho(r')r'^2 dr'\,, \end{equation}

with $\Gamma$ being the Lorentz factor of the outflow.

The trends of the thickest lines can be interpreted as follow. If there
is no opacity, typically at early times, it is because the jet is able
to propagate through and emerge from the remnant before triggering an
external shock. At later time, the opacity increases because the amount
of material at a radius that is larger than the location of the
companion BH grows. Finally, at very late times, the remnant has become
so big that the main effect is the dilution of the density, causing the
opacity to then decrease. It is interesting to notice that the
feature of an increasing and then decreasing optical depth is unique to
this model and would potentially be a way to distinguish it from models
in which the accretion responsible for the late time flares is onto the
same engine that drove the FLBOT itself.

For the case of AT2022tsd, which is especially constraining due to its
fast variability, we also show the opacity for radiation produced by the
disk (or any other source) at the position of the BH. In this case,

\begin{equation} \tau_{\rm{BH}} = \kappa \int_{0}^{\infty} \rho(r') dr'
\label{eq:tau2} \end{equation}

and we use a thinner line with respect to the one used for $\tau$.
Notice how the opacity is larger at early times, since this radiation is
produced at the location of the BH and not at the external shock
radius\footnote{Note that we do not include here a possible wind
component discussed in Sect.~\ref{sec:emission}, which would increase
the opacity for any photon produced close to the BH.}. Finally, again
only for the case of AT2022tsd, we plot the optical depth that a
transient originating from the exploding star would see. This is
computed as 

\begin{equation} \tau_{\rm{SN}} = \kappa \int_{10^{10}
\mathrm{cm}}^{\infty} \rho(r) dr\,. \end{equation}

We notice that the physical lower limit is needed to avoid divergence
($\rho(r)\propto(1/r)$ in this regime) and was chosen to represent the
radius of a compact star. We also emphasize that this integral is over
$r$, the distance from the center of the exploding SN, and not from the
center of the BH as in Eq.~\ref{eq:tau1} and~\ref{eq:tau2}. We notice
that any transient originating at the radius of the exploding star will
likely be diffused, at least for a year after the explosion, and would
not display fast variability as seen in AT2022tsd after about one month.

Finally, short vertical lines of the corresponding color show the time
at which EM activity was observed in the three considered FBOTs. We see
that for the AT2022tsd case the remnant is thin at optical/NIR
frequencies, for both the ES and disk emission. It is, instead, opaque
in X-rays. For the case of AT2020mrf, Figure~\ref{fig:optDepths} shows
that the remnant is thick in X-rays. This is at odds with the
observation of variable X-ray emission from the source. It should be
noted, however, that we have neglected the clumping in our opacity
calculations, and photons might be able to propagate without diffusion
through the thinner medium between clumps, if the covering fraction of
clumps is less than unity. In addition, the X-ray flux from AT2020mrf is
overproduced by the model with the fiducial ejecta mass
$M_{\rm{ej}}=$M$_\odot$ (see central panel of Figure~\ref{fig:emission})
and a lower ejecta mass would both decrease the opacity and improve the
agreement in the predicted and observed flux. We notice, finally, that
it is critical that the radiation is produced by the external shock
(which is the case in the central panel of Figure~\ref{fig:emission}).
Radiation from the disk and from the exploding star would be subject to
a much higher opacity that is unlikely to be reduced enough by
geometrical considerations and/or by reducing the remnant mass. Finally,
 AT2018cow lies at  $\tau\sim1$ in the X-ray band, and does not have an
issue with diffusion.

\section{Population models}\label{sec:populations} To determine the
frequency of binary systems which match the conditions of our model, we
run population synthesis calculations with the rapid binary population
synthesis code COSMIC\footnote{https://cosmic-popsynth.github.io/}
\citep{Breivik2020}. As a baseline, we take a fiducial set of model
assumptions that apply the default prescriptions  in COSMIC for a
population of $5\times10^6$ binaries with metallicity
$Z=0.8\,Z_{\odot}$. We initialize the population with primary masses
$M_1$ following \cite{Kroupa2001}, secondary masses sampled uniformly
between $[0.08\,M_{\odot}, M_1]$, and orbital periods and eccentricities
following the distributions fit by \citet{Sana2012}. The most important
model assumptions for determining the properties of binaries hosting BHs
with stripped-star companions are the BH formation mechanism (and its
resulting natal kick) and the division between which systems undergo
stable mass transfer or common envelop evolution when a star fills its
Roche lobe. We assume that compact objects form using the `delayed'
prescription from \citet{Fryer2012} and that natal kicks are drawn from
a Maxwellian distribution with $\sigma=265\,\rm{km/s}$ where the natal
kick is linearly reduced by the fallback fraction, $f_b$ as defined by
\citet{Fryer2012}. We assume that the outcomes of Roche-overflow
interactions are determined through critical mass ratios, $q_c =
M_{\rm{donor}}/M_{\rm{accretor}}$, as defined by \citet{Neijssel2019}
such that Roche-overflowing systems with $q\geq\,q_c$ enter a common
envelope (CE) and systems with $q<q_c$ enter stable mass transfer (SMT).
In the case of a CE, we assume that the orbital energy is used with
perfect efficiency to eject the envelope ($\alpha=1$ in the
$\alpha-\lambda$ formalism \cite{vandenHeuvel1976}). For SMT, we assume
that accretion is limited to $10$ times the thermal rate
$\dot{M}_{\rm{thermal}} = M_{\rm{accretor}} /\tau_{\rm{thermal}}$. All
mass that is not accreted is assumed to carry away the specific angular
momentum of the accretor.

\begin{figure} \centering
\includegraphics[width=0.8\textwidth]{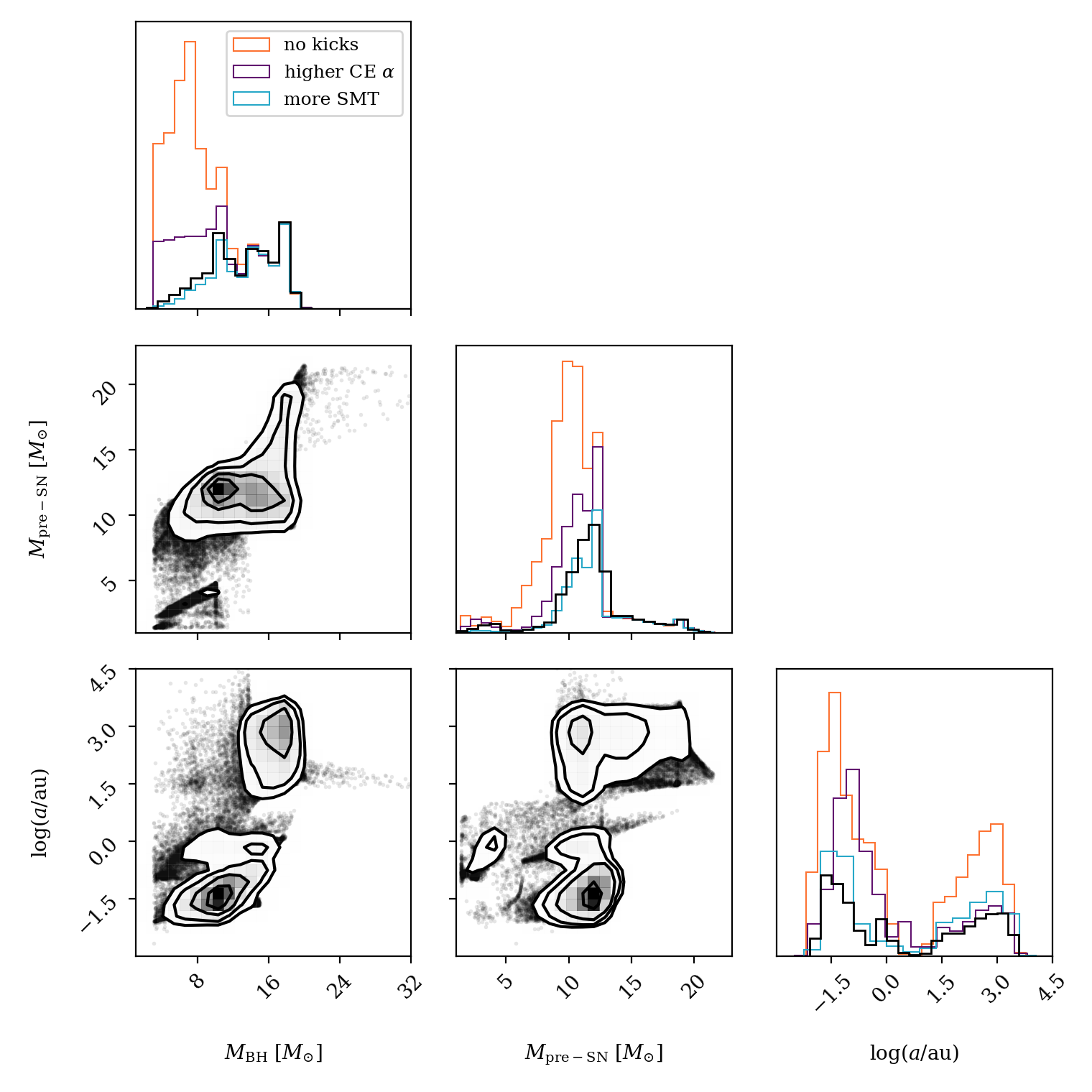} \caption{A
corner plot showing the BH mass, pre-SN stripped star mass, and
separation at the time of the SN for our fiducial model in black. Model
variations are shown in the one-dimensional histograms with different
colors.} \label{fig:pop-corner} \end{figure}

Figure~\ref{fig:pop-corner} shows the orbital properties of binaries
where a stripped-envelope SN occurs with BH companion for our fiducial
model in black. COSMIC assumes that all Roche-overflow mass transfer
circularizes the binary, thus all SNe occur in circular orbits. We find
a significant population of stripped-envelope SNe with BH companions
with $a=10-1000$~AU. These originate from systems which avoid Roche
overflow and contain BHs with masses between $12-18\,M_{\odot}$ and
pre-explosion masses between $10-20\,M_{\odot}$. In orbits with
$a<10$~AU the binary progenitors undergo Roche overflow which leads to
lower BH and pre-explosion stripped star masses. These results suggest
that transients from clumpy accretion onto a BH companion can occur on
both short and long timescales in appreciable quantities (c.f.
Figure~\ref{fig:theThree}.)

To test the sensitivity of our results to the binary evolution
assumptions we made, we ran three model variations. First we assume BHs
receive no natal kick at their formation, but do experience the
symmetric kick due to instantaneous mass loss in the orbit
\citep{Blaauw1961}. Next we consider the case where the CE ejection
efficiency is raised to $\alpha=5$ such that some form of internal
energy in the envelope can be used to unbind the envelope
\citep{Fragos2019}. Finally, we increase the critical mass ratio for CE
to $q_c=3$ consistent with \citet{Belczynski2008} which increases the
number of systems which experience SMT.

Our model variations are shown in the one-dimensional histograms of
Figure~\ref{fig:pop-corner}. Since each population was run with
self-consistent initial conditions, the histogram heights accurately
reflect the relative rates from each variation. The `more SMT' variation
produces the smallest change and leads to fewer close ($a<1$ AU)
binaries. Both the `higher CE $\alpha$' and `no kicks' variations lead
to significantly more systems. In the CE variation, fewer binaries merge
prior to the stripped-envelope SN because there is less shrinking in the
orbit from both the mass transfer phase initiated by the BH progenitor
and the mass transfer phase initiated by the stripped star progenitor
relative to the fiducial model. This effect is more pronounced in close
orbits since these binaries are more likely to survive a natal kick at
the formation of the BH while wider systems will unbind due to the kick.
Finally, when BHs receive no natal kick, the overall population size
significantly increases both in close and wide systems. This effect is
most pronounced in the BH mass panel where the lowest mass BHs are
normally disrupted due to appreciable natal kicks under the fiducial
model.

Finally, we note that for late-time transients ($10-10^3$ days post SN),
the binaries are wide enough to completely avoid Roche-overflow
interactions. These systems are thus free from uncertainties which arise
from mass transfer and are thus an exceptional window into BH formation
and natal kicks. Conversely, future work to study the importance of
ejecta interactions shortly after the supernova may shed important light
on the uncertain binary interactions that produce stripped-envelope SNe
in binaries with close BH companions.\\

\section{Discussion and conclusions}\label{sec:discussion}

We have presented a model for late-time flaring and/or continuous
emission in FLBOTs systems. Our model proposes a scenario in which the
FLBOT is produced by an exploding massive compact star that is part of a
close binary system with a companion BH. In most models discussed in the
literature (e.g. \citealt{Margutti2019,Yao2022,Migliori2024}) the late
emission is explained as activity from the same central engine that
powered the FLBOT. In our model, instead, the late emission is due to
material from the FLBOT ejecta that is accreted onto the companion BH.
This has several advantages. Our model naturally accounts for the delay
between the FLBOT and the late emission, due to the time it takes the
ejecta to reach the orbit of the BH. In addition, since the BH is at the
outer boundary of the ejecta, its emission suffers from much lower
extinction and scattering, allowing for the engine variability timescale
to be preserved. Finally, since the accretion is not driven by fallback
and has an onset time that is different from the explosion time of the
FLBOT, it is natural to observe fast variability like the one seen in
AT2022tsd. To better understand this argument, consider the variability
observed in GRB X-ray flares, which can be understood as due to
late-time engine activity from a central engine accreting clumpy
fallback material \citep{Perna2006,DallOsso2017}. Their variability
timescale obeys $\Delta{t}_{\rm{flare}}/t_{\rm{flare}}\sim0.2$, where
$\Delta{t}_{\rm{flare}}$ is the flare duration and $t_{\rm{flare}}$ the
delay of the X-ray flare photons from the GRB trigger
\citep{Chincarini2010}. For AT2022tsd, minute duration flares were
detected about a month after the FLBOT. These flares have therefore
$\Delta{t}_{\rm{flare}}/t_{\rm{flare}}\sim10^{-5}$, strikingly different
from what observed in GRB X-ray flares. In the model presented here,
$t_{\rm{flare}}$ would not be the time since the FLBOT but rather the
time since the accreting clump has reached the companion BH, which can
be as close as needed to the flare time. We consider the fast
variability observed in AT2022tsd to be the principal reason for why a
model like the one presented here is especially promising.

A better understanding of the physics involved would come from
high-cadence panchromatic follow-up of future events. As shown in
Figure~\ref{fig:emission}, our model explains the fast variable optical
emission as synchrotron from the shocked region between the BH outflow
and the ejecta material. This is very broadband emission and
simultaneous flares in a wide range of frequencies should be observable,
provided that the remnant is thin at all frequencies. In addition, as
the remnant grows in size, opacity is expected to increase and fast
variability should disappear at late times. This is in contrast to
models in which the central engine of the FLBOT and of the flares is the
same, since opacity is expected to be monotonically declining in that
geometry.

Should our model be supported by future data, searching for the
predicted long-term emission (after a quiescent period) following a
FLBOT explosion would be a powerful way to uncover the presence of a
binary companion -- and constrain its orbital separation. This is
especially useful in light of the importance of these systems as
gravitational wave sources.  Additionally, a systematic analysis of the
flare properties can provide valuable, independent constraints on the
intrinsic properties of the SNR clumps. It should also be noted that
this phenomenon may not be unique to FLBOTs. Any system in which an
exploding star has a BH companion in a binary system will likely trigger
accretion and electromagnetic emission (see also
\citealt{Fryer2014,Kimura2017b}). Whether it would be detectable depends
on the system geometry and dynamics, as discussed for the FLBOT case.
Interestingly, our population synthesis modeling shows that binaries
with the right separation should not be uncommon. However, unless the
presence of a BH companion in the tens of AU range causes the stellar
explosion to have the properties of an FLBOT, there is no reason not to
expect fastly variable flares to follow more mundane core-collapse SN
events.

As a final remark, the model that we discussed has some important
simplifications that warrant further analysis. For example, it assumes a
SN remnant with constant density and velocity, characterized by a large
number of identical spherical clumps. In reality, it is more likely that
the remnant is expanding in homologous fashion, with the fastest ejecta
on the leading edge. The equations that we derived for a constant
velocity show that the flares duration and luminosity depend on the
ejecta velocity through the size of the accretion disk and the viscous
timescale. If the clump geometry remains constant through the flow, the
early flares would be expected to be longer and less luminous, with
increased activity at later time. In addition, clumps are likely to have
a distribution in size, shape, and density contrast, favoring a
diversity of flare properties that is not captured by our model in the
simple version presented here.

\section*{Acknowledgments}

R.P. and D.L. thank the organizers of the FEET2024 workshop where this
work was initiated. D.L. acknowledges support from NSF award
AST-1907955. R.P. gratefully acknowledges support by NSF award
AST-2006839.

\bibliography{bibio}

\end{document}